\documentclass{epl}

 \usepackage{graphicx}
 \usepackage[latin1]{inputenc}
 \usepackage{latexsym}
 \usepackage{amstext}
 \usepackage{amsxtra}
 \usepackage{amssymb}
 \usepackage{bm}

 \newcommand{\bs}{\boldsymbol}
 \newcommand{\omegm}{\omega_{\rm mp}}
 \newcommand{\omegd}{\omega_{\rm dp}}

\title{Shape oscillation of a rotating Bose-Einstein condensate}
\shortauthor{S. Stock et al.}
\author{Sabine Stock, Vincent Bretin, Frédéric Chevy and Jean Dalibard}
\institute{ Laboratoire Kastler Brossel\footnote{LKB is a unit\'e
de recherche of Ecole Normale Sup\'erieure and Universit\'e Paris
6, associated to the CNRS.}, 24 rue Lhomond, F-75231 Paris Cedex
05, France }


\date{\today}

\begin{document}

\maketitle
\begin{abstract}
We present a theoretical and experimental analysis of the
transverse monopole mode of a fast rotating Bose-Einstein
condensate. The condensate's rotation frequency is similar to the
trapping frequency and the effective confinement is only ensured
by a weak quartic potential. We show that the non-harmonic
character of the potential has a clear influence on the mode
frequency, thus making the monopole mode a precise tool for the
investigation of the fast rotation regime.
\end{abstract}

The investigation of rotating gases or liquids is a central issue
in the study of superfluids \cite{Donnelly91,Tinkham}. During the
recent years, several experiments using rotating atomic
Bose-Einstein condensates have provided a spectacular illustration
of the notion of quantized vortices
\cite{Matthews99,Madison00,Ketterle1,Hodby01}. Depending on the
rotation frequency of the gas, a single vortex or a regular array
of vortices can be observed experimentally. When the rotation
frequency is increased to a very large value, a new class of
phenomena is predicted, in connection with quantum Hall physics
\cite{Cooper01,Paredes01,Ho01,Regnault03,Fischer02,Read03,Baym03,Baksmaty03}.
For a gas confined in a harmonic potential, the fast rotation
domain corresponds to stirring frequencies $\Omega$ of the order
of the trapping frequency $\omega_\bot$ in the plane perpendicular
to the rotation axis (hereafter denoted $z$). From a classical
point of view, the transverse trapping and centrifugal forces
compensate each other for this stirring frequency, and the motion
of the particles in the $xy$ plane is only driven by Coriolis and
interatomic forces. This situation is similar to that of an
electron gas in a magnetic field, since Lorentz and Coriolis
forces have the same mathematical structure.

In order to approach the regime of fast rotation two paths are
currently being explored. The first approach is implemented in a
pure harmonic potential and is based on evaporative spin-up, i.e.
the selective removal of particles with low angular momentum
\cite{Boulder03}. The stirring frequency $\Omega$ can then be
raised close to $\omega_\perp$ ($\Omega=0.993\,\omega_\perp$ was
reached in \cite{Boulder03b}). The second approach, which is
followed here, consists in adding to the quadratic confinement a
small positive quartic potential, which ensures that the particles
will remain confined even when $\Omega$ exceeds $\omega_\perp$
\cite{Fetter01,Lundh02,Kasamatsu02,Kavoulakis03,Tsubota03,Aftalion03}.
We have recently proven that this method can be successfully
implemented and we have mechanically stirred a rubidium
Bose-Einstein condensate up to $\Omega \simeq 1.05\,\omega_\perp$
\cite{Bretin03}.

All experiments performed so far (including the present one) are
still deeply within the mean field regime, characterized by a
number of vortices $N_v$ well below the number of particles $N$.
Nevertheless in order to prepare for future investigations of
possible quantum-Hall-like states, one must design proper tools of
investigation, such as the study of the eigenmodes of the rotating
gas. Possible examples are the transverse quadrupole modes, which
allow for a measurement of the $z$ component of the angular
momentum \cite{Bretin03}, and the Tkachenko oscillations of the
vortex lattice, recently observed in \cite{Boulder03}. Here we
report on the experimental and theoretical study of the lowest
transverse monopole mode of an ultra-cold gas of rubidium atoms in
the fast rotation regime.

In a trap with axial symmetry along the $z$-axis, excitations can
be characterized by their angular momentum along $z$ (quantum
number $m_z$) \cite{Dalfovo99}. Here we are interested in
excitations carrying no angular momentum ($m_z=0$). In the limit
$\omega_z \ll \omega_\perp$ which is of interest here, it is
possible to identify $m_z=0$ ``transverse monopole" modes, which
mostly affect the atom distribution in the $xy$ plane. For a
non-rotating condensate the lowest frequency for these $m=0$
transverse modes is $\omegm \simeq 2\,\omega_\bot$ (transverse
breathing mode) \cite{Fort00,Chevy02}. In the present work we show
experimentally that the relation $\omegm \simeq 2\,\omega_\bot$
remains valid for large rotation frequencies (as predicted by
\cite{Cozzini03}), as long as the quartic term in the confinement
is not significant. We also explore the region $\Omega\sim
\omega_\perp$, where the quartic term plays an essential role. We
compare the measured monopole frequency with that derived from a
simple hydrodynamic model of an infinite, cylindrical condensate.
Finally we discuss the structure of the mode, which exhibits a
very particular behavior in the range $\Omega \gtrsim
\omega_\perp$.

Our $^{87}$Rb Bose-Einstein condensate contains $3\times 10^5$
atoms. It is produced by radio-frequency evaporation in a
Ioffe-Pritchard magnetic trap, to which we superimpose a far-blue
detuned laser beam. The magnetic trap provides a harmonic
confinement with cylindrical symmetry along the $z$ axis, with
$\omega_z/2\pi=11.0$~Hz and $\omega_\perp^{(0)}/2\pi=75.5$~Hz. The
laser beam adds a negative quadratic and a positive quartic
potential in the $xy$ plane. The quadratic term decreases the trap
frequency $\omega_\perp$ by $\sim 15\%$. The quartic term allows
to explore rotation frequencies around and slightly above the
trapping frequency $\omega_\perp$. It reads $kr^4/4$, with
$r^2=x^2+y^2$ and $k=2.6(3)\times 10^{-11}$~J/m$^4$. The
oscillation frequency of the condensate center-of-mass (dipole
motion) in the $xy$ plane is $\omegd/2\pi=65.6\;(\pm 0.3)$~Hz.
Because of the quartic component of the trapping potential,
$\omegd$ is slightly larger than $\omega_\perp$, even for an
arbitrarily small amplitude of the dipole motion. The condensate
has indeed a finite radius $R_c\simeq 6.5\,\mu$m and therefore
explores regions of space where the contribution of the quartic
term is significant. Using a perturbative treatment of the quartic
term we infer that $\omegd-\omega_\perp\simeq
2kR^2/(7\,m\omega_\perp^2)$, so that
$\omega_\perp/(2\pi)=64.8\;(\pm 0.3)$~Hz.

The procedure for setting the gas in rotation at a frequency
$\Omega\sim \omega_\perp$ has been described in detail in
\cite{Bretin03}. We apply an additional laser beam creating a
rotating anisotropy in the $xy$ plane for two successive phases,
with respective durations of 300~ms and 600~ms. During the first
phase the rotation at frequency $\Omega\sim \omega_\perp/\sqrt 2$
resonantly excites the transverse quadrupole mode, and vortices
subsequently penetrate the condensate. After a 400~ms relaxation
period we apply the second stirring phase at the final desired
frequency. This phase is followed by a 500~ms relaxation period.
By measuring the transverse size of the condensate, we have
checked that it is then rotating at a frequency close to the
stirrer frequency $\Omega$ (within 2\%) for the regime of interest
$\Omega\leq 1.05\,\omega_\bot$. In particular, for $\Omega
\lesssim \omega_\perp$ we observe large vortex arrays involving up
to $\sim 50$ vortices. During the stirring and relaxation periods,
we apply radio frequency (r.f.) evaporative cooling. The r.f. is
set 24~kHz above the value which removes all particles from the
trap. Assuming $\Omega=\omega_\perp$, so that the transverse
confinement is purely quartic, this gives a trap depth $U_0\sim
40\;$nK, hence a temperature $T \lesssim 10\;$nK.

We then excite the transverse monopole mode(s) of the gas by
changing for a period $\tau_0$ the intensity of the laser creating
the quartic potential from $I$ to $I'$. We choose $\tau_0 =2\,$ms,
which is short compared to the oscillation period of the monopole
oscillation $2\pi/\omegm=7.5\;$ms.  We then wait for an adjustable
duration $\tau$ before performing a time-of-flight expansion and
absorption imaging. The images are taken along the rotation axis
$z$ so that we have access to the column density in the $xy$
plane.

\begin{figure}
\centerline{\includegraphics[width=14cm]{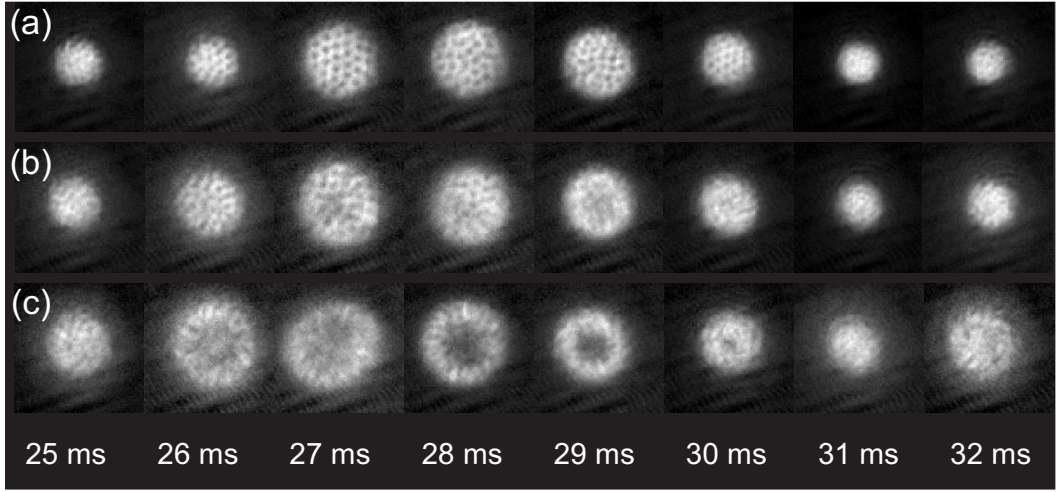}}
\caption{Series of images of the condensate for various waiting
times $\tau$ after excitation. Row a) corresponds to a rotation
frequency $\Omega/(2\pi)= 62$~Hz, row b) to $\Omega/(2\pi)= 64$~Hz
and row c) to $\Omega/(2\pi)= 66$~Hz.} \label{fig:movie1}
\end{figure}

Typical results are shown in figure~\ref{fig:movie1} for
$\Omega/2\pi=62$, $64$, $66$~Hz. Each image corresponds to a
destructive measurement of the atom density after a duration
$\tau=25,\ldots,32$~ms and a 18~ms time-of-flight. In order to
check that the oscillation is in the linear regime, we explored
values between 0 and 4 for the ratio $I'/I$: we found that the
frequency $\omega_0$ is (within experimental uncertainties)
independent of the excitation strength $|I-I'|/I$, and that the
amplitude of the oscillation varies linearly with this strength.
We have also varied the duration of the time-of-flight between 6
and 18~ms. We thus verify that all features reported here are
independent of this duration, which corresponds to a mere scaling
of the initial spatial density distribution.

\begin{figure}
\centerline{\includegraphics[width=14cm]{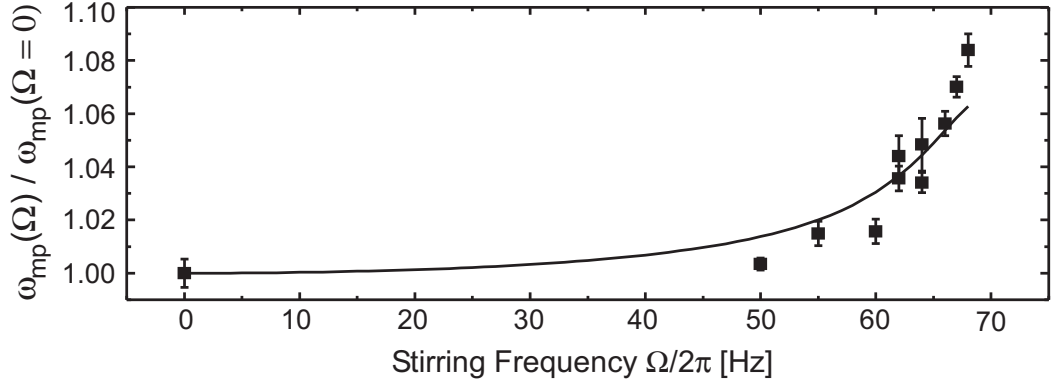}}
\caption{Transverse monopole frequency $\omegm(\Omega)$ normalized
by $\omegm(\Omega=0)$ as a function of the stirring frequency
$\Omega$. For a non rotating condensate ($\Omega=0$) we measure
$\omegm(0)/(2\pi)=130.6$~Hz. The vertical bars indicate the error
given by the fitting routine. } \label{fig:frequency}
\end{figure}

We have taken similar sequences of images for various stirring
frequencies $\Omega$, with $\tau$ varying from 0 to 40~ms by steps
of 1~ms. We extract from these images the transverse size
$R(\tau)$ of the gas. The variations of $R(\tau)$ are well fitted
by a single sinusoidal function, with frequency $\omegm$. The
characteristic lifetime of the excitation is larger than 100~ms,
so that its decay is negligible during the 40~ms period. The
variations of $\omegm$ as a function of $\Omega$ are plotted in
figure~\ref{fig:frequency}. An obvious feature of this plot is
that $\omegm$ varies very weakly for $\Omega$ except in the
vicinity of $\omega_\perp$. We now discuss in more details three
relevant domains of this graph.

(1) For a non rotating gas ($\Omega=0$) we find
$\omegm/(2\pi)=130.6\;(\pm 1.5)$~Hz, which is close to the well
known result $\omegm\simeq 2\omega_\perp=2\pi \times 129.6\,(\pm
0.6)$~Hz \cite{correction}.

(2) When the gas rotates at a frequency $\Omega$ not too close to
$\omega_\bot$, the effect of the quartic potential remains small.
In these conditions we find that the monopole frequency $\omega_0$
stays approximately constant. At first sight this may seem a
surprising result. For a rotating gas in a pure harmonic
potential, the transverse confinement is reduced due to the
centrifugal force. In the calculation of the equilibrium shape of
the gas, the trapping frequency $\omega_\perp$ is thus replaced by
the weaker value $(\omega_\perp^2-\Omega^2)^{1/2}$. One might have
expected that a similar replacement should be done also for the
monopole frequency; this is clearly not the case. For a
cigar-shaped condensate in the hydrodynamic regime, it was proven
in \cite{Cozzini03} that $\omega_0$ stays equal to $2\omega_\perp$
when $\Omega$ varies. In the case of an ideal gas described by
classical mechanics, the equality $\omegm=2\,\omega_\perp$ also
holds for any $\Omega$ as a consequence of the combined action of
centrifugal and Coriolis forces.

(3) When the rotation frequency $\Omega$ approaches
$\omega_\perp$, the contribution of the quartic term becomes more
important and the monopole frequency $\omegm$ deviates
significantly from $2\omega_\perp$. As shown in
figure~\ref{fig:frequency}, this deviation reaches $\sim 8$~\% for
$\Omega/(2\pi)=68$~Hz, i.e. $\Omega/\omega_\bot \simeq1.05$. We
have also plotted in figure~\ref{fig:frequency} the prediction of
a theoretical treatment of the monopole mode in the quartic +
quadratic trap, neglecting the harmonic confinement along the
$z$-direction. The agreement between these predictions and the
experimental data is satisfactory, given the difference in
geometry between the experiment and the model.

The starting point of our theoretical treatment is the set of
equations for rotational hydrodynamics \cite{Cozzini03}, which we
write in the laboratory frame:
 \begin{eqnarray}
\partial_t \rho &=&-\bs \nabla\cdot (\rho \bs v) \label{eq:drhodt}\\
\partial_t \bs v&=& -\bs \nabla ((U+g\rho)/m\,+\,v^2/2)+\bs
v\times(\bs \nabla\times \bs v)\ , \label{eq:dvdt}
 \end{eqnarray}
where $\rho$ and $\bs v$ are the density and velocity field of the
atom distribution. $U=m\omega_\perp^2r^2/2+kr^4/4$ stands for the
trapping potential and $g=4\pi\hbar^2 a/m$ characterizes the
strength of the atomic interactions ($a$ is the scattering
length). As explained in \cite{Sedrakian01,Cozzini03,Chevy03}
these equations are valid when one is interested in a phenomenon
whose characteristic length scale is larger than the distance
between vortices.

The eigenmodes are obtained by linearizing
(\ref{eq:drhodt}-\ref{eq:dvdt}) around the rotating equilibrium
solution
 \begin{equation}
\bs v_{\rm eq}=\bs \Omega\times \bs r
 \qquad
g\rho_{\rm eq}=\mu-U-U_{\rm cen}\ ,
 \label{eq:stationary}
 \end{equation}
where $\mu$ is the chemical potential and $U_{\rm
cen}=-m\Omega^2r^2/2$ the centrifugal potential. We eliminate the
velocity field to get a closed equation for the density variation
$\delta \rho$, assuming an oscillation at frequency $\omega$:
 \begin{equation}
-\bs \nabla \cdot
 \left( \rho_{\rm eq}\,
 \bs \nabla[\delta \rho]
 \right)=
\frac{m}{g}\;(\omega^2-4\Omega^2)\,[\delta \rho]
 \ .
 \label{eq:eigen1}
 \end{equation}
In absence of a quartic term the eigenfrequencies for the $m=0$
modes are
 \begin{equation}
\omega_n^2=4\Omega^2+2n(n+1)(\omega_\perp^2-\Omega^2)\ , \qquad n\
\mbox{positive integer,}
 \label{eq:degener}
 \end{equation}
and the corresponding eigenmodes are polynomials of degree $n$
with respect to $r^2$. For $n=1$, we recover in particular
$\omega_1\equiv\omegm=2\,\omega_\perp$.

When $\Omega$ approaches $\omega_\perp$, all transverse modes
$m=0$ become degenerate. Such a macroscopic degeneracy is
reminiscent of the degeneracy of the energy levels of a single
particle in a uniform magnetic field, leading to the well known
Landau level structure \cite{Landau1}. An equivalent degeneracy
occurs, still at the single particle level, for an isotropic 2D
harmonic oscillator of frequency $\omega_\perp$, considered in a
frame rotating at frequency $\Omega=\omega_\perp$
\cite{noteharmonic2D}. However we emphasize that our result here
holds not for a single particle spectrum, but for the eigenmodes
of a $N-$body system treated in the mean-field approximation. The
occurrence of such a macroscopic degeneracy raises interesting
questions concerning the linear response of the rotating system to
an arbitrary excitation.

When the quartic potential is present, the above degeneracy is
lifted. We set $x=r/R$ in (\ref{eq:eigen1}) with
$R=(4\mu/k)^{1/4}$, to get the dimensionless eigenvalue equation
for the $m=0$ modes:
 \begin{equation}
-\frac{1}{x}\frac{d}{dx} \left( x(1-\epsilon
x^2-x^4)\,\frac{d[\delta \rho]}{dx}
\right)=\Lambda(\epsilon)\,[\delta \rho] \qquad \mbox{where}\qquad
\epsilon=\frac{mR^2}{2\mu}(\omega_\perp^2-\Omega^2)\ .
 \label{eq:eigen2}
 \end{equation}
For each $\Omega$, i.e. for each $\epsilon$, we are interested in
the lowest eigenvalue $\Lambda_0(\epsilon)$ of the hermitian
operator in the left hand side of (\ref{eq:eigen2}). We can then
deduce the frequency of the lowest transverse monopole mode which
is plotted as a continuous line in figure~\ref{fig:frequency}:
 \begin{equation}
\omegm^2=4\Omega^2+\frac{\mu \Lambda_0(\epsilon)}{mR^2}\ .
 \end{equation}
The quantity $\Lambda_0(\epsilon)$ is an increasing function of
$\epsilon$ which we calculate numerically using a variational
method with polynomial trial functions. Just at the critical
rotation $\Omega=\omega_\perp$, $\epsilon=0$ and we get
$\Lambda(0)\simeq 11.5$. The slow rotation limit corresponds to
$\epsilon \gg 1$, in which case $\Lambda(\epsilon) \simeq
8\epsilon$; we then recover the result $\omegm=2\omega_\perp$.

This analysis also explains the periodic apparition of a hole at
the center of the density profile for $\Omega\sim\omega_\perp$, as
observed in figure~\ref{fig:movie1}. The stationary density
profile (\ref{eq:stationary}) varies as $r^{4}$ around the origin,
whereas the mode $\delta\rho$ varies as $r^{2}$ in this region.
The curvature of the density profile is therefore dominated by the
phase of the perturbation.

\begin{figure}
\centerline{\includegraphics[width=14cm]{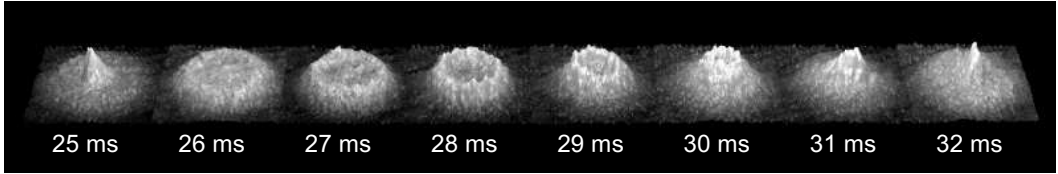}}
\caption{Time evolution of the density profile of the oscillating
cloud, for $\Omega/2\pi=68$~Hz.  } \label{fig:3D}
\end{figure}

For $\Omega\geq \omega_\perp$ and a relatively strong excitation
($I'/I=4$), we observe that the time evolution of the mode
structure becomes asymmetric (see figure~\ref{fig:3D}; this
asymmetry is also slightly visible in the last row of
figure~\ref{fig:movie1}). It consists of a periodic entering
positive density wave, which starts on the edge of the condensate
and gradually moves to the center in a time period $\sim
\pi/\omega_\bot$. A possible interpretation of this unusual
structure is that two (or more) transverse modes $m=0$ are
simultaneously excited with a non-zero relative phase. Since these
modes have similar frequencies (cf. Eq.~(\ref{eq:degener})) the
initial phase difference between them stays nearly constant, and
this can give rise to the observed phenomenon.

To summarize we have presented in this paper a detailed study of
the transverse monopole mode of a fast rotating degenerate Bose
gas. We have shown that the non-harmonic character of the
potential (which is essential for the confinement of the gas) has
a clear influence on the mode frequency. We have also given a
simple analysis of this mode for an infinitely long condensate
which is in good agreement with the experimental data.

Theoretical studies have shown that the addition of a quartic term
in the harmonic potential may lead to the formation of a ``giant
vortex", i.e. a vortex with a circulation larger than the single
quantum $h/m$
\cite{Lundh02,Kasamatsu02,Kavoulakis03,Tsubota03,Aftalion03,Fischer03}.
The large hole appearing at the center of the condensate during
the monopole oscillation when $\Omega \gtrsim \omega_\perp$ should
not be confused with such a giant vortex. We are dealing here with
a transient state of the condensate, while the predicted giant
vortex state is a stationary state of the system, for an
appropriate angular momentum. Another example for a transient
large core in a condensate is provided by an experiment recently
performed in Boulder, in which a hole pierced in a rotating
condensate confined in a purely quadratic trap was shown to
persist for a long time \cite{Engels03}.

We now briefly discuss some perspectives opened by this work.
First we recall that the nature of the gas when the rotation
frequency $\Omega$ is around or above $\omega_\perp$ is still
unclear. The detailed study of \cite{Bretin03} showed that the
number of visible vortices is not sufficient to account for the
measured rotation frequency of the gas. Two classes of explanation
have been proposed to account for this finding. Either the fast
rotating gas cannot be described anymore by a single macroscopic
wave function \cite{Akkermans}, or the vortices are still present
but they are distorted and do not show up clearly in the images of
the condensate. We hope that the present experimental measurement
of the transverse mode frequency can be used to discriminate
between these two hypotheses. In principle a generalization of the
above theoretical treatment to a condensate in a 3D trap is
possible, and the quantitative predictions of such a model could
be compared with the experimental findings.

From the experimental point of view a natural extension of the
present work is to switch to a two-dimension geometry, using a
strong confining potential along the rotation axis. Vortices then
become point objects and the predicted properties of the system
depend on the ratio between the atom number $N$ and the number of
vortices $N_v$. When $N$ is large, the Bose-Einstein condensate
presents a regular vortex array, as already observed in
\cite{Boulder03b}. The array is expected to melt when $N$
decreases to a value of the order of $N_v$ \cite{Sinova02} (see
also \cite{Fischer02}). For $N<N_v$ the quantum Hall regime for
particles should emerge. It is an interesting problem to determine
the signature of these various regimes on the eigenmodes of the
system. For the monopole mode considered here and for a pure
harmonic confinement, it has been predicted in
\cite{Pitaevskii,Kagan96} that the frequency $\omegm$ of the
breathing motion of a 2D gas remains strictly equal to
$2\,\omega_\bot$ for any equilibrium state. This does not hold
anymore in presence of the quartic term, and thus the deviation of
$\omegm$ from $2\,\omega_\perp$ can in principle be used to
monitor the emergence of new quantum phases in the rotating gas.

\acknowledgments We thank S. Stringari and the ENS group for
useful discussions. S. Stock acknowledges support from the
European Union (CQG network HPRN-CT-2000-00125), the
Studienstiftung des deutschen Volkes and the DAAD. This work is
partially supported by CNRS, Coll\`{e}ge de France, R\'egion Ile
de France, and DRED.


\begin{thebibliography}{99}
\bibitem{correction} As for
the dipole mode there is actually a small correction ($\sim 1\%$)
to this relation due to the presence of the quartic potential
$kr^4/4$.
\bibitem{noteharmonic2D}
In a frame rotating at angular velocity $\Omega$ the single
particle Hamiltonian becomes $H'=H-L_z$. Therefore all the levels
of the 2D harmonic oscillator with an angular momentum $m$ equal
to the ecitation number $n$ have an energy
$n\hbar(\omega_\perp-\Omega)$ that goes to zero when
$\Omega=\omega_\perp$.
\end{thebibliography}
\end{document}